\documentclass{article}
\usepackage[utf8]{inputenc}
\usepackage{verbatim, amsfonts, amsmath, amsthm, natbib, har2nat}
\usepackage{graphicx}
\usepackage{url}

\newtheorem{theorem}{Theorem}
\newtheorem{lemma}{Lemma}
\newcommand{\indep}{\perp \!\!\! \perp}

\newcommand{\Ex}[1]{\ensuremath{\mathrm{Ex}}\left({#1}\right)}
\newcommand{\Var}[1]{\ensuremath{\mathrm{Var}}\left({#1}\right)}

\usepackage{tikz}
\usetikzlibrary{shapes.geometric, positioning, calc}
\usetikzlibrary{arrows, arrows.meta}
\tikzstyle{arrow} = [thick, ->,>=stealth]
\tikzstyle{dasharrow} = [thick, dashed, ->,>=stealth]
\tikzstyle{u_vent} = [circle, dashed, text centered, draw=black]
\tikzstyle{o_vent} = [circle, text centered, draw=black]

\title{Constructing valid instrumental variables in generalised linear causal models from directed acyclic graphs}
\author{Øyvind Hoveid\\oyvind.hoveid@nibio.no}
\date{\today}

\begin{document}

\maketitle
\begin{abstract}
%Generalised linear causal models are highly relevant for finding average treatment effects when both treatments and responses are continuous. 
Unlike other techniques of causality inference, the use of valid instrumental variables can deal with unobserved sources of both variable errors, variable omissions, and sampling bias, and still arrive at consistent estimates of average treatment effects. The only problem is to find the valid instruments. Using the definition of Pearl (2009) of valid instrumental variables, a formal condition for validity can be stated for variables in generalised linear causal models. The condition can be applied in two different ways: As a tool for constructing valid instruments, or as a foundation for testing whether an instrument is valid.
When perfectly valid instruments are not found, the squared bias of the IV-estimator induced by an imperfectly valid instrument --- estimated with bootstrapping --- can be added to its empirical variance in a mean-square-error-like reliability measure. 
\end{abstract}

\section{Introduction}
Causal graphs and theoretical results of \citeasnoun{pearl2009causality} have so far not found many applications within empirical econometrics. Based on this fact \citeasnoun{imbens2020potential} suggests that causal graphs are relatively unproductive. An alternative explanation is, though, that causal graph theory never have been accepted as part of the foundations of econometrics. Without taking a final stand in this controversy, I will contribute with improvements to instrumental variables (IV) estimators based Pearl's theoretical concepts.

%Causal inference is required in many contexts. With the currently deep belief in market efficiency is economics nowadays not a prominent area of relevant cases. Questions related to the validity of economic theory and its predictions are nevertheless causal in nature: "What happens if a parameter in policy or nature is modified and everything else is kept constant?"\footnote{I am indebted to Maureen Kilkenny for giving me such a question. Without it would this article not have been written.} 

%Every economist has heard Karl Pearsons comment, "Correlation does not imply causation". But so what? There are several approaches to modelling of causality, propensity score matching, structural equation models, and instrumental variables. In particular the last one is promising and actually developed within econometrics.

The origins of instrumental variables date almost one hundred years back. The geneticist Sewall Wright introduced causal graphs \cite{wright1921correlation}, and his economist father Philip introduced IV, the  IV-estimator, and the equivalent two step least squares estimator\cite{wright1928tariff}\footnote{Based on \citeasnoun{stock2003retrospectives}}. Later is \citeasnoun{haavelmo1943statistical} and \citeasnoun{reiersol1950identifiability} also credited for independent development of instrumental variables in relation to simultaneous equations --- presumably unaware of the contributions of the Wrights.

Instrumental variables have long been recognised as salient ingredients in causal models for non-experimental data. \citeasnoun{wikipedia2020inst} says: \emph{"A valid instrument induces changes in the explanatory variable but has no independent effect on the dependent variable, allowing a researcher to uncover the causal effect of the explanatory variable on the dependent variable."}
However, a statistical test of their validity has so far been lacking within econometrics. This seems due to the fact that validity of an instrument not yet has found a definition with testable implications within this branch of science. 
So far is IV-based causal analysis surviving with the idea that valid instruments are rare and should be selected based on verbal citeria: "detailed institutional knowledge and the careful investigation and quantification of the forces at work in the particular setting" \cite{angrist2001instrumental}.

%As pointed out by \citeasnoun{dufour2003identification} is non-testability closely related to lack of identification. Without precise guidelines for identifying the better instruments, we are also unable to identify the more reliable estimates of causal effect. 

At least for linear models, and for a substantial class of non-linear ones, it is possible to improve on this situation by applying the definition of \citeasnoun{pearl2009causality}: "A variable $z$ is instrumental with respect to a treatment $x$ and an outcome $y$, iff in the causal graph involving $(z,x,y)$, $z$ is a cause of $x$ and $z$ and $y$ are conditionally independent given $x$". The definition needs some translation to point out the testable condition, though, but this will be provided in this article.

A formal test will increase the precision of IV techniques as the selection of instruments may be more precise than when based on the verbal criterion. In addition a test may increase the applicability of IV in two different ways. First, instruments that do not pass the verbal criterion, may pass the test. Secondly, even almost valid instruments may provide informative causal inference. The trick is to add the squared bias due to an almost valid instrument to the variability of the estimated causal effect in a mean-squared-error-like measure.  

Causal modelling with experimental data is well understood since \citeasnoun{fisher1960design}. Basically, when treatments are all under control, % and are manipulated independently, 
the causal effects on responses are found by means of regression models. The main problem with causal analysis and non-experimental data is that treatments have not been controlled. %, nor are varied independently. and need be considered stochastic. %Very few variables except constants rightly deserve the notion of exogenous.    
Of course can control variables be added to the extent they are observed, but there is always a possibility that some unobserved variable make the regression coefficients biased or inconsistent as measures of causal effects. Simpson's paradox --- that an average effect over the non-stratified population has different sign from the corresponding average effects in all population strata --- may even  arise in such situations. One will never know if it did, though, since the clarifying explanation is unobserved. %A third problem is omitted control variables which also may introduce bias for OLS-coefficients. 
Under ideal circumstances, the IV-estimator is expected to solve the problem of unobserved variables affecting regressions.

Actually are perfectly valid instruments not expected to be found. Instrument validity is a property of the population or super-population. Test statistics for a sample of the population, or a population of the super-population, are always affected by sampling errors. A probability need then be assigned to a statement that some variable is a valid instrument in some setting. 

This paper continues in section 2 with a clarification of the close connection between a simple linear causal graph and the corresponding econometric terms. In section 3 is Pearl's definition of a valid instrumental relationship between $z$ and $(x,y)$ translated to econometric equations which should hold at the (super)population level.

Construction of valid instruments, eventual test design and estimation of bias by means of bootstrapping is described in section 4.  
At last section 5 concludes.

\section{Causal graphs translated to econometric terms}
A simple five-noded causal graph as in figure 1 is sufficient to illustrate the relationships between causal graph theory and 
econometrics. It is emphasised that the graph is an hypothesis of the causal relationships --- or equivalently of the data generating process. Alternative hypotheses with arrows pointing in other directions are possible as long as no cycles arise. The graph needs to be a directed acyclic graph (DAG). %Statistical support for one hypotesis does not mean that other plausible hypoteses are less likely. 
Statistics cannot in general say whether $x$ is a cause of $y$ or opposite, or if one hypothesis of causality is more likely than another, as long as there are no missing edges in the graph. Support for the causal hypothesis need be found in theory or common sense. E.g. with reference to figure 1, $w$ might be the external determinants of an economic agent, $x$, might be his capital stocks, and $y$ his short-run production decisions. 

Useful terminology distinguish the starting and ending nodes of an arrow. When nodes $a$ and $b$ is connected with an arrow, $a -> b$, $b$ is named a \emph{child} of $a$, and $a$ is named a \emph{parent} of $b$. Causality runs from parents to children, not the other way. In contrast, correlation runs both ways. In relation to parents and children, the meanings of \emph{ancestors} and \emph{descendants} are the obvious ones.   

This graph assumes that $x$ is a parent of $y$, while $w$ is a parent of both $x$ and $y$. In addition is $\epsilon_x$, representing all unknown parents of $x$, also a parent of $x$, and $\epsilon_y$ is likewise a parent of $y$. Usually are error terms like $\epsilon_x$ and $\epsilon_y$ dropped from the graph, but for sake of explanation they are kept here. 

Each node represents independent observations of a single variable, or a block of variables, with the name of the node. All variables has the same number of observations. If the node has parents, the $i$-th row of a node is a function of the $i$-th rows of the parent nodes. In this case are only linear functions considered and the functions are $x_i = w_i \beta_{w,x} + \epsilon_{i,x}$ and $y_i = w_i \beta_{w,y} + x_i \beta_{x,y} + \epsilon_{i,y}$. %Observe that there is only one error term for each child node --- not two as in state-space models. 

With parameter estimates known, the causal effect on $y$ by changing $x$ to $x'$ is in Pearls terms found with the "do-operator", $y(do(x')) = x'\beta_{x,y}$. The effect is, 
\[ y(do(x')) - y(do(x)) = (x'- x)\beta_{x,y} \]
In more complex cases, the do-operator follows all directed paths from $x$ to  $y$. 

In general Pearl assumes that a probability, $p_i$, is associated with each realisation. $(w_i, x_i, y_i, \epsilon_{i,x}, \epsilon_{i,y})$, and that the error term of a node is independent of the other parent nodes. Here: $\epsilon_{x} \indep w$ and $\epsilon_{y} \indep (w,x)$. Independence is in turn a property of probability distributions $P$, more precisely: $a \indep b$ means 
\[P(a|b) = P(a)\].

Econometrics with linear models tend to skip probabilities and rely on first and second order moments only. The same trick is applicable for causal graphs. Independence is then replaced with weaker condition of orthogonality. As well known, equivalence of orthogonality and independence requires normally distributed variables.

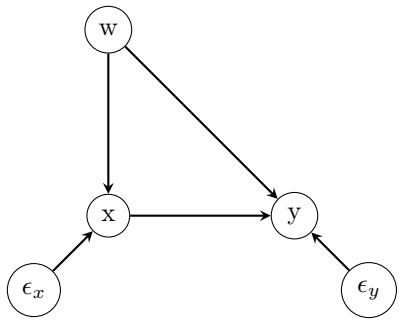
\begin{figure} \label{fig_dag1}
\centering
\begin{tikzpicture}[
	scale=0.7,rotate=45,every node/.style={draw,circle},
	]
\node [o_vent] (y) at (3 * 90 :2.5) {y};
\node [o_vent] (x) at (2 * 90 :2.5) {x} edge[arrow] (y);
\node [o_vent] (w) at (1 * 90 :2.5) {w} edge[arrow] (y) edge[arrow] (x);
%\node [o_vent] (u) at (0 * 90 :2.5) {u} edge[arrow] (y) edge[arrow] (x)  edge[arrow] (w);
\node (eps_y) at (3 * 90 :4.5) {$\epsilon_y$} edge[arrow] (y);
\node (eps_x) at (2 * 90 :4.5) {$\epsilon_x$} edge[arrow] (x);
%\node (eps_w) at (1 * 90 :4.5) {$\epsilon_w$} edge[arrow] (w);
\end{tikzpicture}
\caption{Causal graph of $(w,x,y)$}
\end{figure}
The relevance of the toy-graph of figure \ref{fig_dag1} follows from the following lemma. 
\begin{lemma} \label{l_sort}
 The nodes of an arbitrary directed acyclic graph can be sorted so that each node depends only on parents of lower order 
\end{lemma}
\begin{proof}
Take first nodes without parents in arbitrary ordering. If none exists, the graph contains a cycle. Take then nodes with ordered parents in arbitrary ordering, and continue until there are no more nodes dependent only on the ordered ones. Either there are no more nodes at all, and all nodes are ordered, or there is a subset of nodes depending on at least one non-ordered node. In the latter case the subset contains a cycle. 
\end{proof}
Such sorting is helpful in complex graphs.

For the situation depicted in figure 1 the following lemma will also be proved.
\begin{lemma} \label{l_main}
Let $(w,x,y,\epsilon_x, \epsilon_y)$ be blocks of variables of length $n$ with $(w,x,y)$ non-singular and with zero-means, linearly related as
\begin{equation} \label{x_reg1}
 (w,x,y) 
\begin{pmatrix}
1   &-\beta_{w,x}  &-\beta_{w,y} \\
0   &1          &-\beta_{x,y} \\
0   &           &1
\end{pmatrix} = (w, \epsilon_x, \epsilon_y) 
\end{equation}
with error terms, $(\epsilon_x, \epsilon_y)$, orthogonal to predictors:
\[ w \perp \epsilon_x, \! (w,x) \perp \epsilon_y \]
All relevant variables affecting $x$ and $y$ are observed.

Then:
\begin{itemize}
\item all nodes with no parents are orthogonal \[  w \perp \epsilon_x, \! (w,\epsilon_x) \perp \epsilon_y \]
\item all parameters are equivalent to those of OLS
\item all linear causal effects are given by OLS-parameters
%\item all parameters equivalent to those of a maximum normal likelihood
%\item conditional independence $(w \indep y)|x$ equivalent to $\beta_{w,y} = 0$, and to $w \perp (y - x \beta_{x,y})$.
\end{itemize}
\end{lemma}

\begin{proof}
By assumption $\epsilon_x \perp w$ and $\epsilon_y \perp w$. 
Since, $\epsilon_y \perp x$ is equivalent to $\epsilon_y \perp (\epsilon_x + w\beta_{w,x})$, and $\epsilon_y \perp w$, $\epsilon_y \perp \epsilon_x$. Orthogonal nodes without parents are then proved.

With regard to OLS-parameters, it is well known that OLS parameters imply orthogonality between errors and predictors. To show the opposite, because of assumed orthogonality: 
\[ 0 = (w,x)^T \epsilon_y
= (w,x)^T \left(y - (w,x) 
\begin{pmatrix}
\beta_{w,y} \\ \beta_{x,y}
\end{pmatrix} \right) 
= (w,x)^T y - (w,x)^T (w,x) 
\begin{pmatrix}
\beta_{w,y} \\ \beta_{x,y}
\end{pmatrix} \]

Multiplication with $\left((w,x)^T (w,x)\right)^{-1}$ on both sides, possible by non-singularity, shows that parameters are OLS. 
\[ 
\left((w,x)^T (w,x)\right)^{-1} (w,x)^T y =  
\begin{pmatrix}
\beta_{w,y} \\ \beta_{x,y}
\end{pmatrix} 
\]

There are three blocks of direct causal effects $w->x$, $w->y$ and $x->y$, given by $\beta_{w,x}$, $\beta_{w,y}$ and $\beta_{x,y}$. There is also a total effect of $w$ on $y$, given by $\beta_{w,x}\beta_{x,y} + \beta_{w,y}$, 
\end{proof}

This simple lemma suggests that there is nothing mysterious with the causal graph. Everything is related to standard econometric terms. A slight contrast is that the graph is a system of variables where $x$ plays a double role as both dependent and independent. Such systems are not alien to econometrics either.

On the other hand has econometrics concerns with respect to efficiency.
If errors are heteroscedastic, econometrics prefer generalized least squares (GLS) as opposed to OLS. Clearly, for sake of efficiency, GLS can and should be applied also for causal graphs provided the same weighting matrix is applied for every first and second order moment. 

Econometricians often express concerns over residuals being correlated with observed variables. In the causal graph is global correlation between error and ancestor nodes only arising when an arrow between two nodes is wrongly omitted. In the current simple case that is not the case. Another aspect is local correlation. Both in econometrics and in causal analysis could estimation be made more efficient by transforming the observations to make them closer to normally distributed as in generalised linear models. A particularly important 
3
 case is that of binary variables which can be transformed to normals via the log-odds ratio.

The more important contrast between econometrics and causal graph analysis is related to unobserved confounders to be illustrated with another causal diagram in the next section.

\section{The definition of valid instruments}
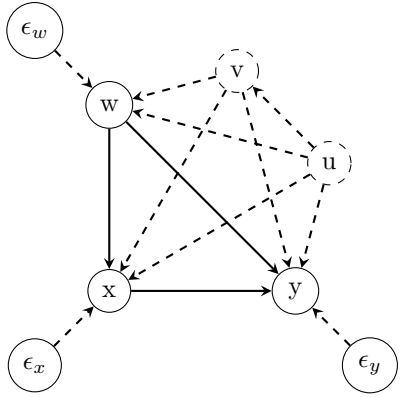
\begin{figure} \label{fig_dag2}
\centering
\begin{tikzpicture}[
	scale=0.7,rotate=45,every node/.style={draw,circle},
	]
\node [o_vent] (y) at (3 * 90 :2.5) {y};
\node [o_vent] (x) at (2 * 90 :2.5) {x} edge[arrow] (y);
\node [o_vent] (w) at (1 * 90 :2.5) {w} edge[arrow] (y) edge[arrow] (x);
\node [u_vent] (v) at (0.33 * 90 :2.5) {v} edge[dasharrow] (y) edge[dasharrow] (x)  edge[dasharrow] (w);
\node [u_vent] (u) at (-0.33 * 90 :2.5) {u} edge[dasharrow] (y) edge[dasharrow] (x) edge[dasharrow] (w) edge[dasharrow] (v);
\node (eps_y) at (3 * 90 :4.5) {$\epsilon_y$} edge[dasharrow] (y);
\node (eps_x) at (2 * 90 :4.5) {$\epsilon_x$} edge[dasharrow] (x);
\node (eps_w) at (1 * 90 :4.5) {$\epsilon_w$} edge[dasharrow] (w);
\end{tikzpicture}
\caption{Causal graph of $(u,v,w,x,y)$}
\end{figure}

In the case portrayed in figure \ref{fig_dag2}, are two blocks of unobserved variables $(u,v)$ added to figure \ref{fig_dag1}. These blocks may contain omitted variables affecting one observed variable at a time, and confounders affecting several simultaneously. There is no loss of generality in this relatively simple structure of unobserved variables. 

The presence of unobserved variables means regression coefficients may be biased. Measurement errors come from single omitted variables, selection bias come from confounders.

Both econometrics and causal graph analysis recognise these two reasons why regressions may turn wrong and instrumental variables may be helpful. In addition econometrics has a third one, simultaneity. As shown in lemma \ref{l_sort}, a model over an acyclic graph has a recursivity that avoids simultaneous equations in the econometric sense. Simultaneity will be further commented in the final section.

Regressions can be done with regard to observable variables as for figure \ref{fig_dag1} with equivalent orthogonality conditions and identical outcome. There is no reason to expect that this outcome provide causal information. However, it will bring information on suitable weighting matrices, $\Omega$, leading to less heteroscedasticity of error terms. It will also bring information on suitable transformations of variables, so that error terms becomes closer to normally distributed. Both model modifications will make later use of IV more efficient.

Clearly, $w$ is not a set of valid instruments as the requirement that $w$ affects $y$ only through $x$ is not satisfied. 

\begin{theorem} \label{t_test1}
 With reference to figure \ref{fig_dag2}, when $(u,v,w,x,y)$ are distributed as multivariate normal with $\Ex{u,v,w,x,y} = 0$, then the following are equivalent
 \begin{itemize}
     \item $w$ and $y$ are independent conditional on $x$
     \item $w$ is orthogonal to the residuals of $y$ OLS-regressed on $x$
     \item $\beta_{w,y} = 0$
 \end{itemize}
\end{theorem}

\begin{proof}
When $(u,v,w,x,y)$ are multivariate normal with $\Ex{u,v,w,x,y} = 0$, so are $(w,x,y)$. Let $\Var{w,x,y} = \Sigma$. The distribution of $(w,y)|x$ is also multivariate normal with expectation,  $x \Sigma_{x^T x}^{-1}\Sigma_{x^T (w,y)}$ and variance, $\Sigma_{(w,y)^T (w,y)} - \Sigma_{(w,y)^T x} \Sigma_{x^T x}^{-1}\Sigma_{x^T (w,y)}$. The independence of $y$ and $w$ conditional on $x$ means the latter matrix is block diagonal. That is:
\[
\begin{split}
0 &= \Sigma_{w^T y} - \Sigma_{w^T x} \Sigma_{x^T x}^{-1}\Sigma_{x^T y}\\
 &= \Ex{w^T y} - \Ex{w^T x} \beta_{x,y}\\ 
 &= \Ex{w^T y} - \Ex{w^T x \beta_{x,y}} \\ 
 &= \Ex{w^T y - w^T x \beta_{x,y}} \\ 
 &= \Ex{w^T \left(y - x \beta_{x,y} \right)}
\end{split}
\]
and $w$ should be orthogonal to the residuals of $y$ OLS-regressed on $x$. 
By the regression anatomy formula, $\beta_{w,y} = 0$.

Equivalence follows from reverse statements.
\end{proof}

\begin{figure} \label{fig_dag3}
\centering
\begin{tikzpicture}[
	scale=0.7,rotate=45,every node/.style={draw,circle},
	]
\node [o_vent] (y) at (3 * 90 :2.5) {y};
\node [o_vent] (x) at (2 * 90 :2.5) {x} edge[arrow] (y);
\node [o_vent] (z) at (1 * 90 :2.5) {z} edge[arrow] (x);
\node [u_vent] (v) at (0.33 * 90 :2.5) {v} edge[dasharrow] (x)  edge[dasharrow] (z);
\node [u_vent] (u) at (-0.33 * 90 :2.5) {u} edge[dasharrow] (y) edge[dasharrow] (x) ;
\node (eps_y) at (3 * 90 :4.5) {$\epsilon_y$} edge[dasharrow] (y);
\node (eps_x) at (2 * 90 :4.5) {$\epsilon_x$} edge[dasharrow] (x);
\node (eps_w) at (1 * 90 :4.5) {$\epsilon_w$} edge[dasharrow] (w);
\end{tikzpicture}
\caption{Causal graph of $(u,v,z,x,y)$ with $z$ as valid instrument}
\end{figure}
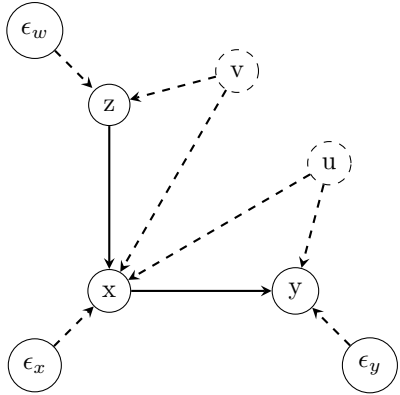

Theorem \ref{t_test1} suggests that a modification of $w$ to $z$, so that $\beta_{z,y} = 0$, make $z$ a set of valid instruments. It is not so that unobserved $u$ vanish, neither that the direct effects of $z$ on $y$ vanish, it is the combination of a direct effect of $z$ on $y$ plus a non-causal correlation through $u$ that vanish by cancelling each other. A relevant diagram is then figure \ref{fig_dag3} where node $w$ is replaced by node $z$ with no direct effect on $y$ and no non-causal correlation between $z$ and $y$.

Parameters $\beta_{z,x}$ and $\beta_{x,y}$ are still without causal content because of expected omissions and confounding, but the IV-estimator, $\beta_{xy}^{IV}$, is expected to be consistent. Actually, it will be proved below that it is, provided variables are normally distributed.

\begin{theorem} \label{t_test2}
 With reference to \ref{fig_dag3}, when $(z,x,y)$ are distributed as multivariate normal, and $z$ and $y$ are independent conditional on $x$, then is the IV-estimator consistent as a measure of the causal effect of $x$ on $y$.
\end{theorem}

\begin{proof}
By lemma \ref{l_main} the variables $(u,v,\epsilon_z,\epsilon_x,\epsilon_y)$ are all orthogonal. Thus $z$, as a linear function of $v$ and $\epsilon_z$ is orthogonal to $u$, and this is also the case with the predictor of $x$, $\widehat{x}(z) = z (z^T z)^{-1} z^T x$.
We now have $x$ separated in three orthogonal parts:
\[ x = \widehat{x}(z) + \epsilon_x + u \]
which affects $y$:
\[ y = \widehat{x}(z) \beta_{\widehat{x},y} + \epsilon_x \beta_{\epsilon_x,y} + u \beta_{u,y} + \epsilon_y\]
By their orthogonality none of these components change when others are omitted. The two latter ones are unobserved and will be omitted. The coefficient $\beta_{\widehat{x},y}$ is then the causal effect of $\widehat{x}$ on $y$ and also the causal effect of $x$ on $y$ for $(u,\epsilon_x)$ kept fixed.

The effect can be specified as:
\[ \beta_{\widehat{x},y} = \left(\widehat{x}^T \widehat{x}\right)^{-1} \widehat{x}^T y
= \left(x^T z (z^T z)^{-1} z^T x \right)^{-1} x^T z (z^T z)^{-1} z^T y\]
When $z^T x$ has an inverse, this expression simplifies to:
\[ \beta_{\widehat{x},y} = \left(z^T x \right)^{-1}  z^T  y\]
which is the well-known IV-estimator.
\end{proof}

It should be observed that normal distributions seem necessary in this situation. With a normal distribution the assumption of conditional independence is equivalent to an orthogonality constraint. Orthogonality without normality is a weaker assumption without direct connection to Pearl's  definition of valid instruments. Possibly will some consistency results still hold, but the efficiency of estimation is expected to be better the closer one gets to normality.

\section{Construction of valid instrumental variables --- and other tricks}
Obviously can theorem \ref{t_test1} be applied to test the validity of some instrumental variable. There is, though, a more constructive application. Valid instruments can be constructed.

Start out from some variables $w$ as in figure \ref{fig_dag2}. Compute also the residuals of $x$ OLS-regressed on $w$, $\eta_{x} = x - w (w^T w)^{-1} w^T x$. Both blocks of variables have causal effect on $x$. A linear combination $z =(w,\eta_{x}) \lambda $ has also this property. Valid instruments should have the property
\[ \lambda^T (w,\eta_{x})^T \left(y - x (x^T x)^{-1} x^T y)\right) = 0\]
When the row dimension of $w$ is at least as large as that of $x$, and that again at least as large of that of $y$, a sufficient number of instruments will most likely be found.

When there is a space of valid instruments, one may even look for those having $\|z_k\| = 1$ and $|z_k^T x|$ large.

When sufficient numbers of valid instruments are not found, one might proceed with the least invalid instruments, $z$, and compute $z'$ satisfying the orthogonality constraints with minimum deviations, $(z-z')_k^T $. Some small deviations are not devastating. After all the orthogonality constraints should hold at the (super-)population levels, not for finite samples.

The bootstrapping technique amounts to make a number of samples, $M$, of size $n$ by random draws with replacement from $(z',z,x,y)$. All estimation routines are repeated for each sample $(z,z',x,y)_m$. The IV-estimators are computed as, $\beta_{m}^{IV}$ with $z$ as instrument and $\beta_{m}^{IV'}$ with $z'$. 

Estimates of the expectation and variance of IV-estimators, given that $z$ is a valid instrument, are found as $\Ex{\beta^{IV}}$ and $\Var{\beta^{IV}}$. If it does not hold, some bias will be involved. An estimate of the bias is $\Ex{\beta^{IV'} - \beta^{IV}}$. The squared bias should be added to the sampling variance, $\Var{\beta^{IV}}$, for a mean-square-error-like reliability measure.

\section{Concluding comments}
With an econometric approach to causal analysis with instrumental variables, one hopes for some natural experiment where $z$ is random and therefore independent of both $u$ and $v$ in figure \ref{fig_dag3}. In that case both $u$ and $v$ and their arrows disappear from the model. In addition one needs to argue that $z$ has no direct effect on $y$.

As shown in this article, the causal graph analyst has more options. He may start out from a vector of other observed variables, $v$, with a causal effect on $x$, and find a set of valid instruments satisfying Pearl's definition, which the econometrician would not think of. 

For some reason has the econometric community, and to considerable extent also the statistical, not embraced causal graphs. An inherent conceptual problem with most causal graphs is their non-uniqueness. Statistical methods cannot in general decide the direction of causal arrows. In social sciences, as opposed to physics, are precise theories not known, and it is not at all obvious how causal graphs should be drawn. Therefore may several causal graphs or DGPs be equally valid for a set of observations. 

This should be no problem for the case of instruments and causality. The hypothetical direction of causality from $z$ to $x$ and from $x$ to $y$ is based on other evidence than statistics. At least, the block sorting of variables here in $w$, $x$ and $y$ may be less controversial than a complete sorting of single variables.  

\citeasnoun{heckman2013causal} see causal graphs based on DAGs as a straight-jacket. They would like to see simultaneity with causality going both ways also treated with graphs. After all, it was with simultaneous equations instrumental variables first entered econometrics as a tool of identification \cite{haavelmo1943statistical}. 

The obvious counterargument is that tools tailored to specific situations may be more productive than universal ones. It has actually been shown in this article that the restriction to acyclic models open for a much larger space of valid instrumental variables than what econometricans are able to find with their informal analysis of each case.

To some extent, has simultaneity already been considered in the analysis here. Variable block, $y$, consists possibly of a set of simultaneous variables with no predefined causal order between members. Assumed causality is only coming from the blocks $x$ or $w$. Bringing that into account, there is a conditional distribution, $P(y|x,w)$, which in some sense also defines a sort of \emph{simultaneous causality}. with a partition of $y$ into $(y_1,y_2)$, can conditional distributions $P(y_2|y_1,x,w)$ also be formed, and an effect of $y_1$ on $y_2$ conditional on $x$ and $w$ can be defined as: 
\[ \partial_{y_1} \Ex{y_2 P(y_2|y_1,x,w)} \]
Simultaneous causality is not directional and is a different concept than graphical causality, though.

In addition, it should be remembered that acyclic graphical models actually can cope with simultaneity issues within temporary contexts. With an auto-regressive model
\[ y_t = A y_{t-1} + \epsilon_t \]
and a matrix of auto-regressive parameters, $A$, causality may flow both ways at the same time.

Time will show whether such solutions will satisfy grumpy econometricians.

\bibliography{inc}
%\printbibliography

\end{document}